# Application of the nnU-Net for automatic segmentation of lung lesion on CT images, and implication on radiomic models

**Authors:**

Matteo Ferrante[a,*,^], Lisa Rinaldi[b,*], Francesca Botta[a], Xiaobin Hu[c], Andreas Dolp[d], Marta Minotti[e,^], Francesca De Piano[e,^], Gianluigi Funicelli[e,^], Stefania Volpe[f,g], Federica Bellerba[h], Paolo De Marco[a], Sara Raimondi[h], Stefania Rizzo[i,k], Kuangyu Shi[d,k], Marta Cremonesi[b], Barbara A. Jereczek-Fossa[f,g], Lorenzo Spaggiari[g,l], Filippo De Marinis[m], Roberto Orecchia[e,n], Daniela Origgi[a]

[a] Medical Physics Unit, IEO European Institute of Oncology IRCCS, via Ripamonti 435, 20141 Milan, Italy

[b] Radiation Research Unit, IEO European Institute of Oncology IRCCS, via Ripamonti 435, 20141 Milan, Italy

[c] Department of Informatics, Technical University of Munich, Arcisstraße 21, 80333 Munich, Germany

[d] Chair for Computer-Aided Medical Procedures, Department of Informatics, Arcisstraße 21, Technical University of Munich, 80333 Munich, Germany

[e] Division of Radiology, IEO, European Institute of Oncology IRCCS, via Ripamonti 435, 20141 Milan, Italy

[f] Division of Radiation Oncology, IEO European Institute of Oncology IRCCS, via Ripamonti 435, 20141 Milan, Italy

[g] Department of Oncology and Hemato-Oncology, University of Milan, Via Festa del Perdono 7, 20122 Milan, Italy

[h] Department of Experimental Oncology, IEO European Institute of Oncology IRCCS, via Ripamonti 435, 20141 Milan, Italy

[i] Clinica di Radiologia EOC, Istituto Imaging della Svizzera Italiana (IIMSI), via Tesserete 46, 6900 Lugano, Switzerland.

[j] Università della Svizzera italiana, via G. Buffi 13, 6900 Lugano, Switzerland

[k] Department of Nuclear Medicine, Bern University Hospital, University of Bern, Freiburgstrasse 18, 3010 Bern, Switzerland

[l] Division of Thoracic Surgery, IEO, European Institute of Oncology IRCCS, via Ripamonti 435, 20141 Milan, Italy

[m] Division of Thoracic Oncology, IEO, European Institute of Oncology IRCCS, via Ripamonti 435, 20141 Milan, Italy

[n] Scientific Direction, IEO, European Institute of Oncology IRCCS, via Ripamonti 435, 20141 Milan, Italy

^ affiliation at the time of the study

**Corresponding author: Botta Francesca (francesca.botta@ieo.it)**

* These authors contributed equally to the work.

**Abstract**

Lesion segmentation is a crucial step of the radiomic workflow. Manual segmentation requires long execution time and is prone to variability, impairing the realisation of radiomic studies and their robustness.

In this study, a deep-learning automatic segmentation method was applied on computed tomography images of non-small-cell lung cancer patients. The use of manual vs automatic segmentation in the performance of survival radiomic models was assessed, as well.

**METHODS**

A total of 899 NSCLC patients were included (2 proprietary: A and B, 1 public datasets: C).

Automatic segmentation of lung lesions was performed by training a previously developed architecture, the nnU-Net, including 2D, 3D and cascade approaches.

The quality of automatic segmentation was evaluated with DICE coefficient, considering manual contours as reference. The impact of automatic segmentation on the performance of a radiomic model for patient survival was explored by extracting radiomic hand-crafted and deep-learning features from manual and automatic contours of dataset A, and feeding different machine learning algorithms to classify survival above/below median. Models' accuracies were assessed and compared.

**RESULTS**

The best agreement between automatic and manual contours (DICE=0.78±0.12) was achieved averaging predictions from 2D and 3D models, and applying a post-processing technique extracting the maximum connected component.

No statistical differences were observed in the performances of survival models when using manual or automatic contours, hand-crafted or deep features. The best classifier showed an accuracy between 0.65 and 0.78.

**CONCLUSION**

The promising role of nnU-Net for automatic segmentation of lung lesions was confirmed, dramatically reducing the time-consuming physicians' workload without impairing the accuracy of survival predictive models based on radiomics.

**Abbreviations**

CT  = computed tomography

NSCLC = Non-small-Cell Lung Cancer

VOI = volume of interest

HU = Hounsfield unit

ML = machine learning

RF = Random Forest

SVM = Support Vector Machine

MLP = Multilayer Perceptron models

Std = standard deviation

## 1. Introduction

In Europe, lung cancer is the second most common malignancy in men and the third most common in women with higher incidence rates in developed over undeveloped countries. Non-small cell lung cancer (NSCLC) accounts for 80-90% of lung malignancies, and includes adenocarcinoma, squamous cell carcinoma and large cell carcinoma (Planchard et al., 2018; Sung et al., 2021). After radiological diagnosis, the treatment of these malignancies can involve different clinical pathways.

Advanced methods can be applied to the radiological images, especially computed tomography (CT), to derive useful information for the physicians in order to improve diagnostic accuracy and choose the best treatment for each patient (Forghani et al., 2019; Ibrahim et al., 2021).

In recent studies, radiomic analysis of medical images has shown promising results for the prediction of therapy outcomes, survival probabilities and other clinical endpoints, including i.e. tumour type, stage, mutation status, and presence of metastasis (Huang et al., 2016; Zhang et al., 2017; Botta et al., 2020; Ninatti et al., 2020; Cucchiara et al., 2021). In addition, Artificial Intelligence (AI)-based methodologies have been increasingly used, either within the radiomic workflow or alone (Hosny et al., 2018a; Coudray et al., 2018; Xu et al., 2019; Lakshmanaprabu et al., 2019; Avanzo et al., 2020; Binczyk et al., 2021; Jiao et al., 2022). The final aim of these approaches is to mine high-level information from radiological images, which are not visible to the human eye but might be relevant for clinical purpose.

The traditional radiomic approach is based on the calculation of descriptors, named *radiomic features*, from the numerical content of the medical images. These descriptors quantify different properties of the area of the image under investigation, among which shape, signal intensity and texture, and can be analysed for possible association with clinical endpoints using different methodologies, including statistical approaches or machine learning AI techniques (Gillies et al., 2016). In the most advanced AI applications, such descriptors can be learnt by deep learning algorithms rather than being calculated with hand-crafted tools (Hosny et al., 2018b; Afshar et al., 2019; Castiglioni et al., 2021; Papadimitroulas et al., 2021).

One of the main bottlenecks of radiomic studies is the segmentation of the volume of interest (VOI), typically the lesion in oncological applications. This is a crucial step, since all subsequent operations of the radiomic workflow (feature extraction and model development) are concerned with the VOI only. Segmentation is most often performed manually by one or more expert physicians, and it is a very demanding task, especially for being extremely time consuming. Moreover, the delineation of the lesion is prone to intra- and inter-reader variability (Parmar et al., 2014; Pavic et al., 2018; Owens et al., 2018; Joskowicz et al., 2019; Bianconi et al., 2021a). AI can play a role in this context as well, with the introduction of neural networks (Siddique et al., 2021; Yu et al., 2022). Once conveniently trained and validated, such algorithms can drastically reduce the time spent by radiologists in monitoring and segmenting lesions and at the same time would allow to increase the number – and therefore the reliability – of the datasets under investigation.

The introduction of automatic segmentation has been investigated in recent years also for the contouring of lung tumours (Kido et al., 2020; Binczyk et al., 2021; Liu et al., 2021; Bianconi et al., 2021b).

The present study fits into this scenario with the following two purposes. The first aim was to train a state-of-the-art model for automatic segmentation, the nnU-Net (Isensee et al., 2021), on CT images of patients affected by different stage lung cancer. Different configurations were implemented and compared, aiming to maximize the accuracy of the automatic segmentation in comparison to the manual segmentation performed by physicians, taken as reference standard. The second aim was to assess to what extent the use of automatic segmentation in replacement of manual contouring affected the performance of a clinical predictive model. To this purpose, radiomic models for the prediction of five-years survival were built on a clinical population of early-stage NSCLC patients, comparing the performances obtained when extracting hand-crafted or deep radiomic features from manual vs automatic segmentation.

## 2. Material and methods

### 2.1 Patient population and CT acquisition

In this study three different datasets of patients diagnosed with NSCLC were analysed, two retrospectively collected at the European Institute of Oncology (Milano, Italy) - Dataset A and Dataset B - and one publicly available, Dataset C.

More in detail, Dataset A included 270 patients staged up to pT3pN1M0, undergoing surgery soon after acquisition of a diagnostic contrast-enhanced CT at the European Institute of Oncology without pre-operative chemotherapy. The clinical characteristics of this population have been described elsewhere (Botta et al., 2020). The information on the acquisition parameters of the CT image can be found in the **Supplementary Materials - Table S2**.

Dataset B included 217 patients extracted from a database of 261 patients affected by advanced NSCLC, undergoing chemotherapy after the acquisition of a diagnostic contrast-enhanced CT image. Clinical characteristics of this population are reported in **Supplementary Materials - Table S1**, while CT acquisition information are listed in **Supplementary Materials - Table S2.** Only patients undergoing CT imaging at the European Institute of Oncology were selected.

The Institutional Review Board of the European Institute of Oncology approved this retrospective study (UID-2078), waiving the need for informed consent.

Dataset C included 412 patients extracted from the *Lung1* public dataset (Aerts et al., 2019) of the NSCLC-Radiomics collection on The Cancer Imaging Archive platform (TCIA) (Clark et al., 2013). The full dataset is composed of 422 patients, of these, 10 were excluded due to metadata instability or corrupted labels. Clinical information can be retrieved from (Aerts et al., 2014) and the TCIA website (https://wiki.cancerimagingarchive.net/display/Public/NSCLC-Radiomics).

### 2.2 Segmentation

#### 2.2.1 Manual segmentation

One lesion per patient was segmented on the axial CT images by manually delineating the border of the lesion slice by slice. Datasets A and B were contoured by three radiologists with different degree of experience, after a common agreement on the procedure. The common criteria among radiologists included the adoption of both the lung and the mediastinal visualisation windows (width of 1500 HU and level of -600 HU and width of 350 HU and level of 40 HU, respectively) according to the lesion location and to the contrast with the surrounding tissues; moreover, vessels were excluded and opacity along the lesion edges were included. Segmentation was performed on the AWServer platform (v. 3.2 Ext. 2.0 tool, GE Healthcare) and saved in RT Structure format.

Original segmentations from Dataset C segmentations were downloaded from the online archive and imported on 3D Slicer version 4.10.2 (Fedorov et al., 2013). Each segmentation was revised by a single radiation oncologist, who edited the volume of interest, as needed. Specifically, all nodal areas were excluded from the gross tumour volume to overcome possible inconsistencies between features extracted from the primary tumour and regional lymph nodes. Blood vessels were excluded, as well.

The final masks were then saved in nearly raw raster data (nrrd) format.

#### 2.2.2 Automatic segmentation: training and testing

The nnU-Net, previously developed for automatic segmentation tasks in medical imaging (Isensee et al., 2021), was adopted for the present study. The net is trained by providing as input a set of images (training set) along with the corresponding manual segmentation performed by the physicians, considered the reference standard to be learnt. During the training phase the parameters of the network are learnt in order to produce automatic segmentations as much similar as possible to the manual reference segmentation. Then, in inference, it is tested on previously unseen images (test set) for which the automatic segmentation is provided as output.

The model training and testing was repeated in twelve different configurations to investigate the variation of the model performance with different source data and network architecture. The configurations were chosen by using different combinations of the three datasets (A, B and C) as training or test sets, and varying the spatial resolution of the images and of the manual segmentation masks. The split between training and test sets was performed randomly, without overlapping between them to avoid overfitting. This procedure aimed at increasing the intrinsic variability of the tested datasets, and identifying the model yielding the best performance in terms of computing power, training timing and segmentation performance. Images and masks were used either at full resolution (512×512) or reduced to half spacing (256×256) using the SimpleITK Python library for voxel resampling (Yaniv et al., 2018).

In **Table 1** all the configurations investigated in this study are listed along with the description of the training modality, the initial image resolution, and the number of cases used for training and testing.

**Table 1:** Investigated combinations of the three datasets for the training and testing of nnU-Net.

| # | Configuration ^ | | Training modality | Initial image resolution | # pts training | # pts testing |
|---|---|---|---|---|---|---|
| | Training | Testing | | | | |
| 1 | A | A * | ensemble(2D, 3D fullres) | 256×256 | 220 | 50 |
| 2 | A + B | A + B * | ensemble(2D, 3D fullres) | 256×256 | 296 | 128 |
| 3 | A + B | A + B * | ensemble(cascade, 3dfullres) | 512×512 | 340 | 147 |
| 4 | C | C * | ensemble(2D, 3D fullres) | 512×512 | 328 | 84 |
| 5 | C | A | ensemble(2D, 3D fullres) | 512×512 | 328 | 79 |
| 6 | C | B | ensemble(2D, 3D fullres) | 512×512 | 328 | 66 |
| 7 | C | A + B | ensemble(2D, 3D fullres) | 512×512 | 328 | 147 |
| 8 | A + B + C | A | ensemble(2D, 3D fullres) | 512×512 | 668 | 80 |
| 9 | A + B + C | B | ensemble(2D, 3D fullres) | 512×512 | 668 | 67 |
| 10 | A + B + C | C | ensemble(2D, 3D fullres) | 512×512 | 668 | 84 |
| 11 | A + B + C | A + B + C * | ensemble(2D, 3D fullres) | 512×512 | 668 | 231 |
| 12 | B + C | A | ensemble(2D, 3D fullres) | 512×512 | 629 | 270 |

^ the two 'configuration' columns, one for the training and the other for the test sets, report the name of the dataset/datasets from which the images were extracted.

* for this configuration the training and test sets were created starting from the same dataset/datasets, but the patients used for the two sets were not the same, meaning that there was no overlapping between the two groups.

Regarding training modality, at least two different models were trained for each configuration, and the performance of the ensemble model was evaluated. The ensemble configuration combines the outputs of multiple architectures by averaging, for each voxel, the probabilities predicted by the networks of belonging to the lesion instead of the background. Except for configuration #3, an ensemble(2D, 3D fullres) approach was used, which combines a 2D and a full resolution (fullres) 3D U-Net architecture. In the 2D architecture the training is performed by taking as input each slice separately and using a 2D convolutional kernel. In the 3D architecture, the CT images are analysed considering also the adjacent slices with 3D kernels in order to catch the volumetric (inter-slices) information and thus provide the learning process with the maximum context. In configuration #3, instead, a cascade approach was applied. The cascade architecture consists first of all in a 3D U-Net trained using down-sampled inputs; the outputs of the segmentation are then up-resampled to the original size and used as additional inputs to another 3D U-Net, which is trained at full resolution. This approach was tested in one configuration only, due to the intensive requirements in terms of computational resources.

In all cases the network was trained using the nnU-Net 'all' flag, except in the case of the 'cascade' model. For this last, the training was repeated five times by operating a five-fold cross validation. This allowed to test numerous configurations in acceptable times by measuring the variations of the performance.

More information about the nnU-Net parameters and available architecture modalities can be found in the **Supplementary Materials** (*Additional information on the segmentation network (nnU-Net)*).

A customised post-processing algorithm — different from the one proposed in the original nnU-Net paper — was developed and applied to the output of the network. In its original version, the nnU-Net package applies a post-processing, returning only the largest connected component with a non-maxima suppression algorithm. We calculated, instead, the connected components in three dimensions using the Python connected-components 3D library (https://pypi.org/project/connected-components-3d/#description) to individuate and separate the segmentation of multiple lesion or false positives. In particular we separated each connected volume estimated as a lesion and computed its volume. Estimates of the ensemble and 3D approaches were proposed with decreasing confidence in order of volume size, so that it was possible to choose the most suitable segmentation, even when multiple lesions were identified.

A pipeline of the segmentation procedure is shown in **Figure 1**.

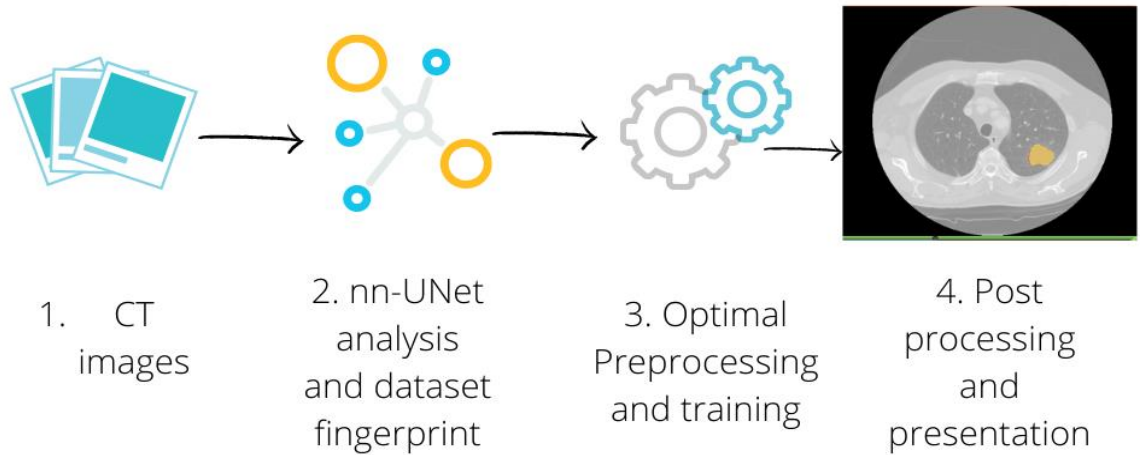

**Figure 1 Segmentation Pipeline**. The CT images of the patients with a lung lesion were collected, along with the manual contours used as the ground truth (1). For each configuration we separated the cases into training and test set. The training set was passed through the nnU-Net framework (2), which automatically adapted the pre-processing to perform the optimal training (3). We finally applied the post-processing to the best performing model to present separately each connected component provided by the network (4).

The segmentation network was trained for a minimum of 400 epochs, defined as 250 batch size as in the original nnU-Net paper on a server with a Nvidia RTX 2080 Ti GPU Card with 11 GB of dedicated RAM memory. Each training modality required from 10 to 20 hours to be completed.

### *2.2.3 Automatic segmentation performance*

The performance of the automatic segmentation pipeline on each testing datasets was assessed by computing the DICE coefficient, a measure of the degree of volume overlapping between the manual (ground truth) and the automatic segmentation (Rizwan I Haque and Neubert, 2020). It ranges between 0, when the two segmentation masks do not share common voxels, and 1, when they are perfectly overlapped. In addition, the percentage of lesions correctly identified by the network was assessed by calculating the number of lesions with a DICE greater than 0, greater than 0.5 and greater than 0.8 out of the total cases in the test set.

## 2.3 Survival prediction

Radiomic models to classify patients according to survival were performed on dataset A, extracting radiomic features from either manual or automatic segmentation, with hand-crafted and deep-learning methodologies, as detailed in the next paragraphs. To this purpose, the automatic contours generated by training the nnU-Net on datasets B + C and testing on dataset A were adopted (configuration #12 in **Table 1**). The reason for choosing this configuration was related to the availability of survival follow-up data for patients in dataset A.

Two additional steps were included in the segmentation pipeline: first, among all the identified connected components, the one with the highest DICE, meaning the highest overlapping with the ground truth, was selected. Then, all the cases for which the DICE coefficient was lower than 0.3 were excluded from the analysis (Haarburger et al., 2020). These two steps, applied taking advantage of the availability of the ground truth manual segmentation for all patients, were intended to simulate the role of the physician in a real clinical application of the automatic algorithm, where manual segmentation is not available. In this case, the physician is expected to revise the segmentations proposed by the automatic algorithm, rejecting the cases for which the segmented object was not the lesion (step 1) or the lesion was correctly identified by the automatic algorithm, but contoured in a wrong way (step 2). For this reason, hereinafter we will refer to this refinement operations as "radiologist simulation" procedures.

### 2.3.1 Hand-crafted radiomic features extraction

Hand-crafted features were extracted from both the manual and the automatic masks using the open-source tool Pyradiomics (v. 2.2.0) (van Griethuysen et al., 2017). The features were extracted from the 'original' images, meaning that no filter was applied to the CT image before the extraction. Moreover, we analysed only features extracted in 2D from each axial slice and averaged among all the slices of the mask, as recommended when the voxels are not isotropic (https://arxiv.org/pdf/1612.07003.pdf).

As pre-processing techniques, pixel size in the axial plane was resampled ('sitkBSpline' interpolation, default in Pyradiomics), and the grey-level intensities were discretised to a fixed bin width of 25 HU (Hounsfield Units).

The categories of features analysed in this study were: *Shape*, *First Order, Gray Level Co-occurrence Matrix* (*glcm*), *Gray Level Run Length Matrix* (*glrlm*), *Gray Level Size Zone Matrix* (*glszm*), *Neighbouring Gray Tone Difference Matrix* (*ngtdm*) and *Gray Level Dependence Matrix* (*gldm*).

### 2.3.2 Deep learning feature extraction

A pipeline involving a neural network was developed to select and extract features autonomously in order to compare or integrate the traditional hand-crafted radiomic approach with a deep one.

Due to the limited size of dataset A, a transfer learning approach was adopted, relying on models pre-trained for more general purposes.

More in detail, a pre-processing pipeline was developed using the Medical Open Network for AI library (MONAI) (MONAI Consortium, 2020), first resampling the image at a fixed size of 1×1 mm in the axial plane and 1.5 mm in the perpendicular plane (spline interpolation), and then cutting the image to a fixed size of 224×224×152, centred at the centre of the segmented volume, in order to be compatible with the chosen network architecture. Subsequently, the ACSConv library (Yang et al., 2021), based on PyTorch (Paszke et al., 2017, 2019), was used to transform a pre-trained 2D model into a 3D model. Basically, the 2D convolutional layers were replaced by their three-dimensional versions by processing the 2D weights to derive the corresponding 3D version, resulting in a three-dimensional pre-trained model, particularly suitable in medical images for being focused on the processing of axial, coronal and sagittal views.

The ResNet152 model (He et al., 2016) was finally applied to extract 2048 deep features from the latest AdaptiveAvgPool3d layer of the PyTorch implementation.

### 2.3.3 *Survival model implementation*

Survival characteristics of the dataset A population, updated at October 2021, were collected. In particular, for each patient, the survival expressed in months after the date of CT examination was recorded. Based on this information, patients were dichotomized according to 5-year survival (Goldstraw et al., 2016).

Different machine learning (ML) models were then trained in order to predict the survival class (below/above 5-years) based on hand-crafted and/or deep radiomic features, extracted from either manual or automatic contours. This analysis allowed to investigate to what extent the use of manual or automatic segmentation can affect the performance of this kind of radiomic-based dichotomous survival model.

A preliminary analysis was performed on the training dataset, comparing different ML algorithms and different hyperparameters combinations, with the purpose of identifying the algorithm and hyperparameters with the best model performance (cross-validation). The investigated algorithms were: Random Forest (RF), Support Vector Machine (SVM) and Multilayer Perceptron models (MLP). For each model, 50 hyperparameter combinations were considered. To this purpose, the Weight & Biases (W&B) (Biewald, 2020) Python library was used. This optimisation phase was performed considering only features extracted from manual segmentations, which represent the gold standard segmentation in this study, and taking hand-crafted, deep and hybrid features (the latter being a combination of hand-crafted and deep features obtained by concatenating these two groups of features) separately.

Subsequently, once the best model and the best combination of hyperparameters were identified according to accuracy metric, the selected model was trained for all the different cases under analysis, using in input hand-crafted features, deep features or hybrid features, extracted from manual or automatic contours.

Finally, a ten-fold cross validation was performed on the entire dataset in order to obtain a statistic measure of the accuracy. The results of the ten-fold cross validation were compared using t-test to check if the distributions were statistically different.

In this way, the impact on model performance of using manual vs automatic segmentation and hand-crafted radiomic features vs deep features vs hybrid features was assessed.

The pipeline for the survival model implementation is schematised in **Figure 2**.

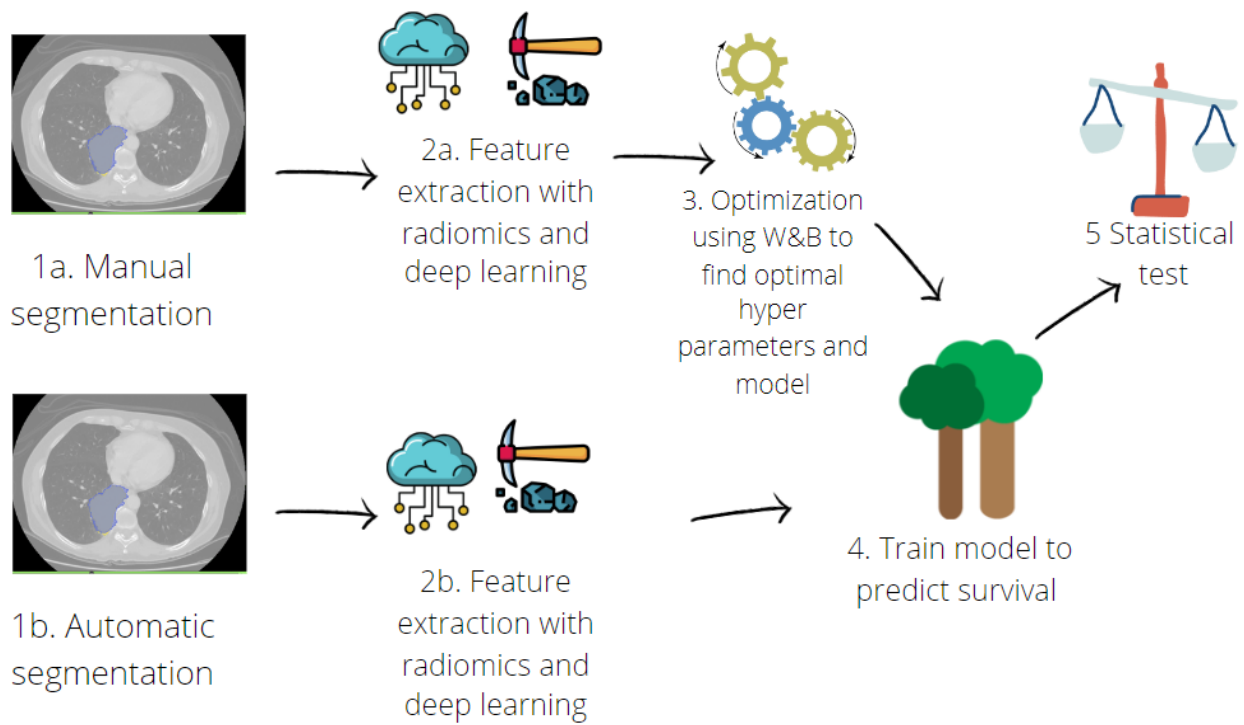

**Figure 2**: **Pipeline for predictive model implementation**. Hand-crafted and deep features were extracted (2a) from manual contours (1a). Then, an optimisation step was performed (3) with a random search over hyperparameters of three different machine learning algorithms: Random Forest, Support Vector Machine and Multilayer Perceptron. Once the best model was found, deep learning and radiomic features were extracted also from contours obtained with automatic segmentation (1b, 2b). The predictive model was trained (4) on features extracted from both manual and automatic segmentation, and ten-fold cross validation was performed. Finally, performances were compared with t-test (5).

The entire code, including the segmentation tool, was developed with Python v 3.8 using popular deep learning and machine learning libraries, including PyTorch (Paszke et al., 2017, 2019), MONAI (MONAI Consortium, 2020) and scikit-learn (Pedregosa et al., 2011).

## 3. Results
### 3.1 Manual and automatic segmentation

According to manual segmentation, mean lesion volume was 38.5 cm$^3$ (range 0.2-511.9 cm$^3$) for dataset A, 57.2 cm$^3$ (range 0.1-708.3 cm$^3$) for dataset B, and 68.2 cm$^3$ (range 0.5-648.9 cm$^3$) for dataset C.

The performance of the automatic segmentation pipeline without applying any post-processing and considering only the outputs of the network with a DICE > 0 is shown in **Table 2** for each configuration investigated (**Table 1**). Mean DICE coefficient (± standard deviation) and percentage of lesions correctly identified by the automatic tool (DICE>0) are reported, along with the percentage of lesions with a good (> 0.50) and excellent (> 0.80) DICE value. The network correctly identified (DICE>0) from 82% up to 94% of lesions, depending on the configuration, with an average DICE among all configurations equal to 0.70 (min = 0.65, max = 0.77). For the majority of the lesions the automatic contours achieved a DICE value larger than 0.50 in all configurations (range: 68-82 %); a DICE over 0.80 was obtained for less than 50% of the lesions in almost all configurations.

The inclusion of the dataset C in the pipeline reduced the segmentation performance in terms of percentage of correctly identified lesions. However, the performance in terms of average DICE was similar to the configurations without dataset C. On the other hand, adding a completely independent cohort — as dataset C — allowed to increase the intrinsic variability of the training set, thus the generalisability of the network.

**Table 2** Results of segmentation, before post-processing and including only the outputs with a DICE > 0. DICE is intended as mean ± standard deviation. The last three columns correspond to the number of lesions with a DICE > 0, DICE > 0.5 and DICE > 0.80, respectively, over the total number of test cases.

| # | Configuration ^ Training | Configuration ^ Testing | DICE | # correctly identified (DICE>0) lesions (%) | # lesions with DICE > 0.50 (%) | # lesions with DICE > 0.80 (%) |
|---|---|---|---|---|---|---|
| 1 | A | A * | 0.65 ± 0.29 | 94 % | 74 % | 38 % |
| 2 | A + B | A + B * | 0.74 ± 0.28 | 93 % | 82 % | 51 % |
| 3 | A + B | A + B * | 0.66 ± 0.32 | 93 % | 71 % | 41 % |
| 4 | C | C * | 0.68 ± 0.33 | 86 % | 71 % | 32 % |
| 5 | C | A | 0.69 ± 0.33 | 83 % | 69 % | 35 % |
| 6 | C | B | 0.71 ± 0.32 | 85 % | 73 % | 40 % |
| 7 | C | A + B | 0.70 ± 0.33 | 84 % | 71 % | 37 % |
| 8 | A + B + C | A | 0.71 ± 0.32 | 88 % | 70 % | 48 % |
| 9 | A + B + C | B | 0.77 ± 0.31 | 87 % | 79 % | 48 % |
| 10 | A + B + C | C | 0.67 ± 0.32 | 83 % | 68 % | 31 % |
| 11 | A + B + C | A + B + C * | 0.71 ± 0.32 | 86 % | 72 % | 42 % |
| 12 | B + C | A | 0.72 ± 0.29 | 91 % | 78 % | 45 % |

^ the two 'configuration' columns, one for the training and the other for the test sets, report the name of the dataset/datasets from which the images were extracted.

* for this configuration the training and test sets were created starting from the same dataset/datasets, but the patients used for the two sets were not the same, meaning that there was no overlapping between the two groups.

Two examples of the output of the segmentation pipeline are reported in **Figure 3,** with automatic segmentation displayed in red, superimposed to the manual segmentation (ground truth) displayed in yellow. **Figure 3a** shows a case with a low DICE (0.43), meaning a not-optimal overlap between the automatic segmentation and the manual one. **Figure 3b** is instead an example of excellent DICE (0.88).

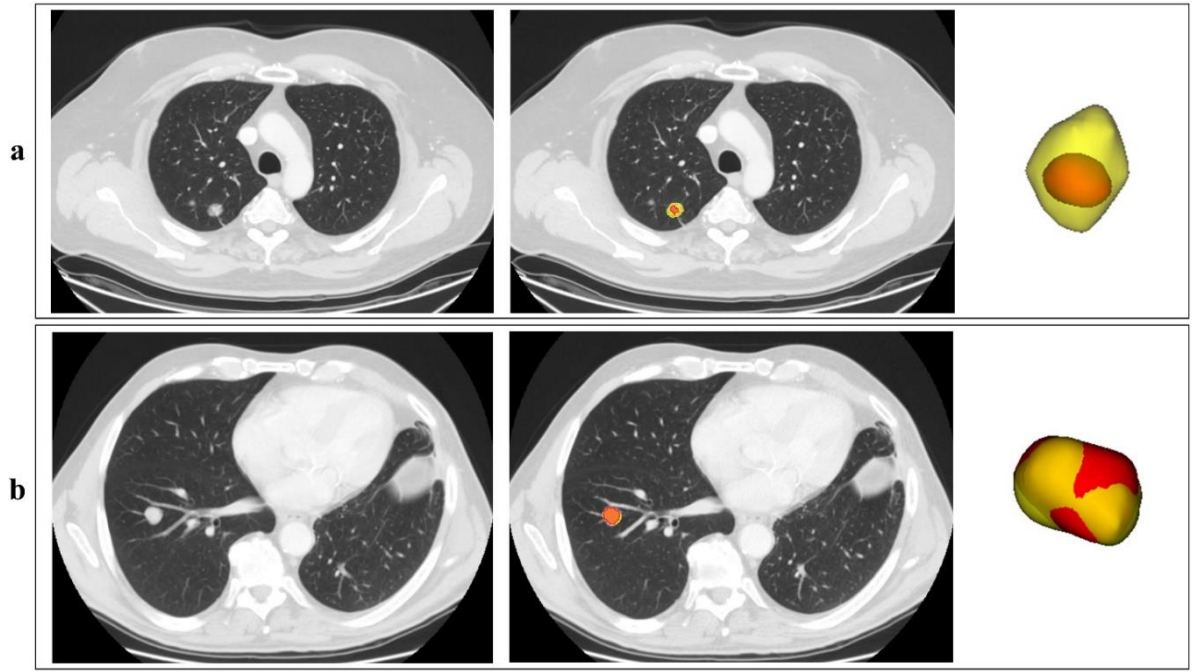

**Figure 3. Examples of the automatic segmentation outputs.** In the picture two outputs of the segmentation network, corresponding to two different patients (a, b), are reported in red, superimposed to the manual segmentation in yellow. For each patient, one axial CT slice including the lesion is reported, without (left) and with (right) indication of the lesion contours; the 3D visual representation of the contours is also reported. Despite the two lesions are quite similar in shape and volume, lesion for patient in picture 2a is characterized by a low-density edge which is not captured by the automatic algorithm.

In case of configuration #12 (training on B+C dataset, testing on dataset A, used for the subsequent survival analysis), when adding the post-processing and the "radiologist simulation" procedure, an improvement of DICE coefficient was observed: the 90% (242/270) of the lesions had a DICE coefficient greater than 0.3, with a mean DICE equal to $0.78 \pm 0.12$ (compared to $0.72 \pm 0.29$ without any post-processing). The distribution of the DICE coefficient over the dataset A in this configuration is shown in **Figure 4**.

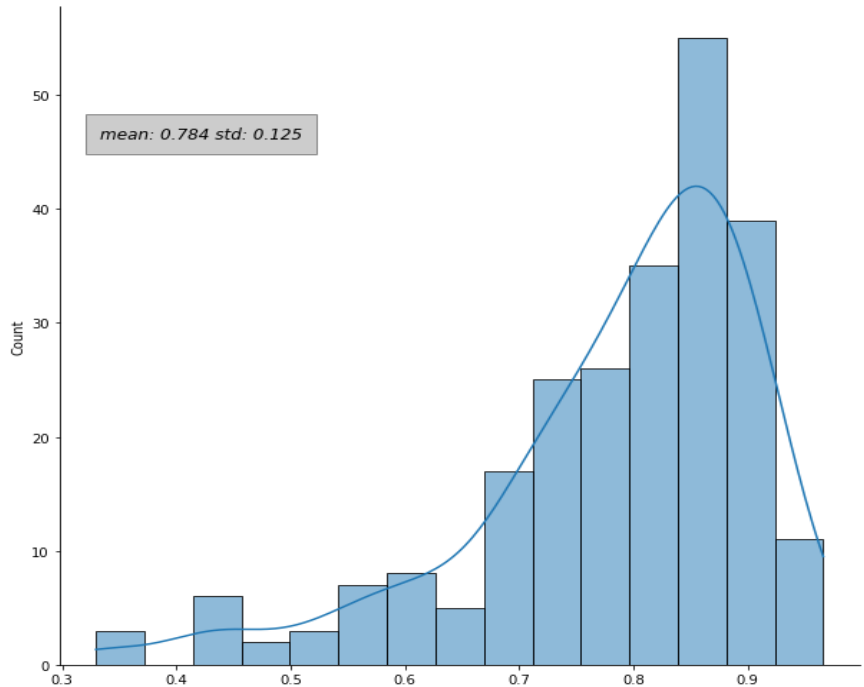

**Figure 4**: **Distribution of the DICE coefficient**. The DICE coefficients were evaluated for the patients in dataset A comparing the automatic and the manual contours in configuration #12. The plot refers to the results after the application of the post-processing and the *radiologist simulation* steps.

To further check the reliability of the segmentation, the volume of the predicted segmentation to the ground truth were compared by computing the relative difference $V_{pred}/V_{real}$;. a mean value of 1.04 was obtained, indicating that in average the predictions were just 4% apart in terms of volume respect of the real ones.

### 3.2 Survival model

When the post-processing and the "radiologist simulation" procedure were applied to configuration #12, 242 patients from dataset A had a valid segmentation and could be used for the prediction of the survival. A total of 132 patients had a survival longer than 5 years, and the remaining 110 fell in the class with survival below 5 years.

Using the W&B library to keep trace of the dependence of the hyperparameters on the three different models tested, the best performances were achieved by using the RF. During the optimisation phase, the *n_estimates* (number of trees in the RF model) and a low value of *ccp_alpha* (regularisation parameter for the cost-complexity pruning algorithm) were positively correlated with accuracy in the training set. As a result, a RF classifier with 1000 trees and a ccp_alpha of 0.01 was selected for the implementation of the classification model.

The model was trained six times in six different cases according to the type of features (handcrafted/deep/hybrid) and to the type of contour used for feature calculation.

In **Table 3**, the accuracy of the model achieved using hand-crafted features, deep features, and a combination of them (hybrid) is reported, separately for features calculated on manual or automatic contours.

**Table 3** Accuracy results of the survival prediction using the RF model, comparing manual and automatic contours. The model performances are given for the three types of features extracted: hand-crafted features (extracted with conventional radiomics), deep learning features (extracted with a deep learning algorithm), and hybrid features (obtained by concatenating the first two types).

| Modality | Manual | Automatic |
| --- | --- | --- |
| **Hand-crafted features** | 0.73 | 0.78 |
| **Deep features** | 0.65 | 0.78 |
| **Hybrid features** | 0.70 | 0.78 |

The p-values of the t-test comparing the accuracies obtained from 10-fold cross validation on the entire dataset are reported in **Figure 5**. No statistical difference was observed (all p>0.05) between the model accuracy obtained with hand-crafted, deep or hybrid features, for both manual and automatic contours, nor when comparing manual vs automatic contours for a given group of features (hand-crafted, deep or hybrid).

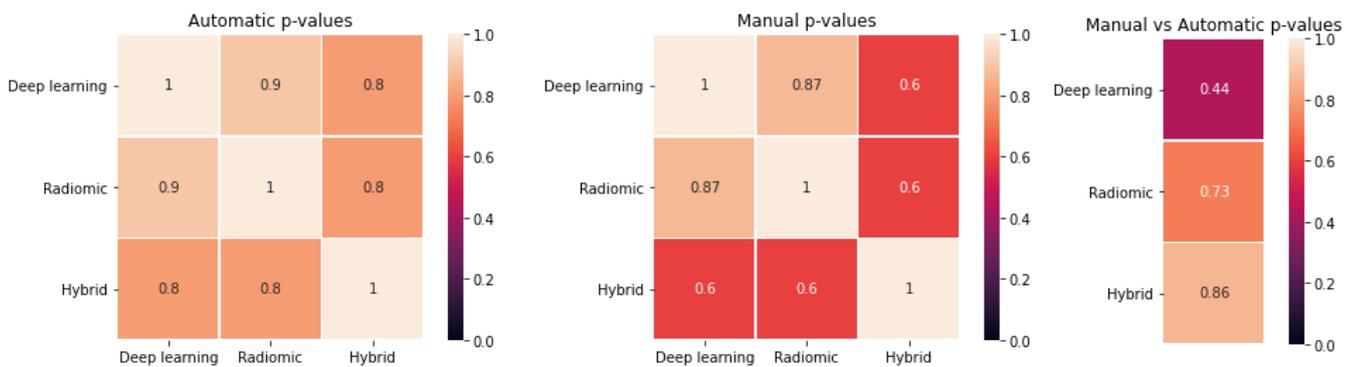

**Figure 5 T-test results of the comparison among the different approaches**. **Left**: p-values for deep learning vs hand-crafted vs hybrid features for automatic segmentation. **Centre**: p-values for deep learning vs hand-crafted vs hybrid features for manual segmentation. **Right**: p-values for manual vs automatic in case of deep learning, hand-crafted and hybrid features.

## 4. Discussion

In recent years, many studies and methodologies emerged with the aim of extracting quantitative information from medical images, potentially relevant to predict clinical outcomes. In the near future, some of these might prove their usefulness to simplify and guide choices towards personalised treatments. However, to be efficiently integrated into the clinical practice, other than robust and generalisable they need to be easily implementable on a routine basis.

Segmentation is in most cases an unavoidable and crucial step of these procedures. Automatic algorithms capable of producing reliable segmentations in a short time could drastically reduce the time required for clinicians for this type of activity, allowing access to much larger segmented datasets during the model development phase. In this regard, it might be not so necessary to train automatic algorithms segmenting the region of interest with equal accuracy as the manual gold standard, rather, the identification of algorithms segmenting the region of interest with enough accuracy to produce comparable predictive models as those obtained with manual segmentation might equally serve the purpose.

In this study a pipeline based on a state-of-the-art framework, the nnU-Net, was developed for automatic segmentation of NSCLC tumour using CT images, testing different network architectures on different dataset configuration. The ability to perform a pre-processing aimed and suited to the characteristics of a CT thoracic image dataset was exploited, and a post-processing algorithm was optimised for the lung lesion segmentation. The automatic contours were compared with the manual ones, drawn by experienced physicians. A survival predictive model was then built, based on radiomic features calculated from either manual or automatic segmentations, and the model performances were compared. In addition, we investigated the difference in model performance, for the clinical outcome under investigation, when using hand-crafted radiomic features, deep-learning features, or a hybrid approach.

Concerning the automatic segmentation procedure, good results were obtained both for the volume overlap between the ground truth and the output of the network (evaluated with the DICE coefficient) and for the number of correctly detected lesions. The average DICE among all the dataset configurations was equal to 0.70 and the percentage of correctly identified lesions was higher than 82% in all the investigated cases (**Table 2**).

These results are quite in agreement with others reported in literature. Gun et al. (Gan et al., 2021) created a combination of a 3D and 2D network to segment lung lesions using a total of 260 patients from a private dataset, achieving a mean DICE coefficient of $0.72 \pm 0.10$. Yang et al. (Yang et al., 2021) developed an ACS (axial-coronal-sagittal) network and tested it on the LIDC-IDRI dataset of lung nodules (Armato et al., 2011). The main idea behind this architecture was to use 2D kernels on the three views separately and then combine them to obtain a 3D output. The best DICE was obtained on the ACS pretrained model architecture, and it was equal to 0.76, outperforming both the 3D (best DICE equal to 0.75) and 2D (DICE equal to 0.69) pretrained networks. Better results were obtained by Haarburger et al. (Haarburger et al., 2020). In this study the authors applied a probabilistic segmentation algorithm based on a 2D U-Net, named PHiSeg network (Baumgartner et al., 2019), on three datasets of different tumours (lung, kidney and liver lesions) and gave a DICE metric of 0.85 (IQR between 0.77 and 0.89) on the LIDC-IDRI lung dataset.

As previously anticipated, a perfect agreement between different segmentations (DICE equal to 1) might not be fundamental for the purpose of predictive model creation. An automatic segmentation not entirely matching the manual gold standard contour but leading to predictive models with comparable accuracy as those built on manual contours reaches the scope of overcoming the manual contouring limitations without impairing the final goal for the clinics. We tested this for a specific clinical endpoint, namely the classification of patients according to survival.

To this purpose, one of the configurations analysed in the segmentation pipeline was selected, the one using dataset B and C for training and dataset A for testing.

As a first step, in this specific configuration, the customised post-processing was applied to the masks obtained from the segmentation network, improving the DICE coefficient from $0.72 \pm 0.29$ to $0.78 \pm 0.12$. The predictive models obtained with hand-crafted, deep or hybrid features calculated from manual or automatic segmentations had comparable accuracy (**Table 3**). Despite the models built on automatic contours provided slightly higher accuracy than those obtained on manual contours (0.78 vs 0.65-0.73), and deep features appeared to perform worse than hand-crafted features in the manual setting (0.65 vs 0.73), a non-significant difference was found (t-test p-values > 0.05) between the performances obtained during ten-fold cross validation with the two types of segmentation and the three categories of extracted features (**Figure 5**).

This was a relevant result proving that a perfect agreement between manual and automatic contours might not be necessary for the purpose of radiomic-based predictive model creation. This might be related to the fact that the disagreement between manual and automatic contours occurs mostly on the lesion edge, which might contribute to a lower extent to the value of radiomic features, the most important lesion characteristics being captured by the internal voxels.

However, it must be pointed out that this result was obtained for the specific clinical endpoint considered in this study (dichotomous classification of survival) and it was tested on a single, multicentre sample. Generalisability to other clinical outcomes, or other populations, should be properly investigated on dedicated datasets. In addition, multiple physicians were involved in the contouring of the lung lesions for the three datasets used in the segmentation pipeline. Even if common criteria were established among them, some variability may have been introduced, reducing the segmentation performance of the network. Further studies are mandatory for the evaluation of the inter-reader variability in manual segmentation. Among other limitations, a relatively small number of images was used for the development of the survival model. Larger datasets might allow us to obtain superior performance. However, in this study we did not focus on the validation of a survival model, but our primary goal was the comparison of the model performance in different scenarios (segmentation modality and/or feature type) when the same group of patients was involved.

## Conclusion

An automatic tool for lung tumour segmentation in CT images was adopted based on the nnU-Net framework and properly adapted with customised post-processing. After testing different network architectures on multiple datasets, the best model achieved an average DICE coefficient of $0.78 \pm 0.12$ after the application of as hoc post-processing technnniques, correctly finding and segmenting the 90% of the tested lesions.

The radiomic features extracted from the so-obtained automatic contours resulted in survival predictive models having comparable accuracy to the ones obtained extracting features from the reference manual contours (accuracy statistically not distinguishable). In addition, hand-crafted and deep radiomic features provided comparable results in terms of predictive accuracy with both segmentation modalities.

These findings support the idea that segmentation tools based on deep learning can be effectively included in the image analysis workflow, dramatically reducing the physician's workload without impairing the accuracy in comparison to the use of manual segmentation. If confirmed, this could simplify the access to large datasets and accelerate the identification of reliable tools and their translation to the clinical practice.

## Acknowledgments

The work was partially supported by the Italian Ministry of Health with Ricerca Corrente and 5x1000 funds. LR, MM and GF were supported by a research grant from the Italian Ministry of Health (GR-2016-02362050). LR was also supported by Fondazione IEO – Radiomic project. SV received a research fellowship from Accuray Inc. SV was partially supported by the Italian Ministry of Health with *Progetto di Eccellenza.*

SV and FBe are PhD students within the European School of Molecular Medicine (SEMM), Milan, Italy.